\documentstyle[aasms4,12pt]{article}
\begin{document}
\title{Short Term Optical Variability in Broad Absorption Line QSOs}
\author{G.C.Anupama and Arati Chokshi}
\affil{Indian Institute of Astrophysics, Bangalore 560 034, India}
\date{}

\begin{abstract}
We present the first results from a pilot programme to monitor the 
short term optical variability in Broad Absorption Line system QSOs. 
Intra-night optical variations of $\sim 5 \%$ were detected on time scales
of $\sim$ one hour in QSOs 0846+156 and 0856+172. Further, the mean
magnitude level decreased in the two objects by $\sim 0.05$ and $\sim 0.15$
magnitude respectively during the
period of our observations. The observed light curves are quite similar to those 
previously seen in the flat spectrum radio-loud sources, especially the BL Lacertae
objects, and can provide important 
constraints for the origin of microvariability, and also a possible evolutionary 
link between the radio-loud and the radio-quiet QSOs.

\end{abstract}
\keywords{quasars: absorption lines --- quasars: general --- quasars: photometry
--- quasars: individual (0846+156, 0856+172)}

\section{Introduction}
Intraday flux variations are a well known characteristic common to BL Lacertae
objects, optically violently variable quasars and flat spectrum radio-loud 
quasars (Wagner \& Witzel 1995). Theoretical explanations for these variations 
invoke relativistic shocks propagating down a jet and interacting with 
irregularities in the flow (Qian et al 1991; Marscher, Gear \& Travis 1992), or 
numerous flares or hotspots on the surface of the accretion disk
believed to surround the central engine (Mangalam \& Wiita 1993). Rapid 
variations in radio-quiet quasars were not established until searches for optical
intra-night variabilities pioneered by Gopal-Krishna, Sagar \& Wiita (1993) 
in an effort to constrain both the models for the origin of the 
variability and also the nature of the nuclear energy source in AGNs. Clear 
detection of low-amplitude optical microvariability were reported for a sample 
of bright radio-quiet  QSOs (Gopal-Krishna, Sagar \& Wiita 1995; Sagar, 
Gopal-Krishna \& Wiita 1996), 
favouring their origin in accretion disks around central massive blackholes.

We have begun a complementary program to study the intra-night optical variability
in QSOs with broad absorption lines (BAL) to probe the 
influence of the BAL phenomena in the microvariabilty within the systems.
Since the BAL systems belong primarily to the sample of radio-quiet QSOs,
and form a minor subset of it, a similarity in their short term light curves would
reflect their origin from the same parent population. However, departures in 
their microvaribility behaviour from those of the parent population would be 
suggestive of a possible link to the BAL phenomena, and its origin. 

In this paper we present the results of our monitoring study for two
BAL QSOs 0846+156 and 0856+172 chosen for our pilot programme, using the 2.3m
Vainu Bappu Telescope.

Section 2 presents the observations, the results are described in section 3;
section 4 discusses our results.

\section{Observations and Analysis}

The BAL QSOs in the present study were chosen from the Hewitt \& Burbidge (1987)
catalog. The observations were carried out at the Vainu Bappu Observatory, Kavalur,
using a Tek $1024\times 1024$ CCD detector at the prime focus of the 2.3m Vainu
Bappu Telescope. Each CCD pixel corresponds to $0.6 \times 0.6$ arcsec$^2$ and the
total sky area covered by the detector is about $10 \times 10$~arcmin$^2$.
The EPADU for the system is 5.6 and read noise about 12 electrons.

We report here the observations of QSO 0846+156 ($z=2.928;\,V=18.3$) and 
QSO 0856+172 ($z=2.311;\,V=19.0$), observed on 1997 February 3 and 4. 
Both objects were observed in the $V$ band.
Several exposures, each of 600s integration time, were obtained for both the
objects over a total period of $\ge 2$~hours. The sky conditions were 
non-photometric on both the nights and the seeing $\sim 2-2.5$~arcsec. 
The data were reduced using the IRAF software package. 
All frames were bias subtracted using a mean bias value. Flat-field correction 
was applied using a median twilight sky flat. Aperture photometry of all the stars
in the field was performed using concentric
apertures of diameter 4, 6, 8, 12, 14 and 18 arcsec centered on the object. Sky
background was subtracted using values measured in an annulus of 3 arcsec at a
radius of 10 arcsec from the center. Based on a growth curve for the brighter
stars in the frame, the 14 arcsec aperture was chosen. Differential magnitudes
were obtained following the method of Howell, Mitchell \& Warnock (1988) and 
Howell (1992), choosing a bright star in the field as the comparison
and a nearby check star with magnitude similar to that of the QSO. The QSO and
the check star being faint, the magnitude measured from the 14 arcsec aperture
is severely affected by the background and read noise. Hence, we use the
magnitude from the 4 arcsec aperture, correcting for the photons from the source
outside this aperture (upto 14 arcsec) using a correction factor estimated based 
on the brighter stars.

\section{Results}
\subsection{0846+156}

This source was observed on 1997 February 3 for $\sim 2$ hours, with a signal-to-noise
ratio of 50--80. The differential magnitudes with respect of the comparison
star (qso-c$_1$) are shown in Figure 1. Also shown in the figure are the differential
magnitudes of the comparison and check star (c$_1$-c$_2$). The scatter in the
individual measurements shown in the figure were estimated following Howell 
et al (1988) and Howell (1992). The QSO light curve reveals a 
variability which could be periodic. There also appears to be a
gradual decline in the mean magnitude level, estimated as a mean line through the
observed points, by about 0.05 magnitude. This trend is not seen in the light 
curve of the check star, which shows an overall variation of $\le 0.01$~mag, similar
to the scatter in the measurements.
We have tested for intrinsic variability in the QSO following the statistical 
method of Howell et al (1988), and find instrinsic variability in the source at 
a 95\% confidence level.

\subsection{0856+172}

This source was observed on 1997 February 4, also for nearly 3 hours, with a signal-to-noise
ratio of 30--50. In Figure 2 we plot the differential magnitudes, along with the
scatter in the individual measurements. This source
also reveals a nearly periodic variability with a period $\approx 40$ min, and a
gradual decline in the mean level by 0.15 magnitude. There is a sudden dip in the 
light curve by 1 magnitude at UT 20.5. There is a slow drift in the (c$_1$-c$_2$) 
light curve by 0.04 magnitude, while no drastic change is seen at UT 20.5.
The statistical test indicates intrinsic variability at a 95\% confidence level.

\section{Discussion}

Optical variability in BAL QSOs has been established on timescales of the order
of years. In his review paper Turnshek (1988) notes that the optical variability
of BAL QSOs differs markedly from that of the non-BAL radio-quiet sample.
Furthermore on these longer timescales, time variability has been observed in
the strength of the BAL features (Turnshek 1988, Barlow et al 1992, Hamann, Barlow \& Junkkarinen 
1997) by as much as a factor of few over timescales of $\sim 0.3$ year (Barlow 
et al 1992, Hamann et al 1997).
Intra-night variability has not been reported since the previous studies were more
suited for detecting long term variations. The observations reported here were
specifically designed to look for the shorter time scale microvariability,
limited only by the S/N consideration at the VBT. 

Intra-night variability is clearly detected in both the BAL QSOs presented here.
0846+156 shows variability with a maxmimum peak-to-peak variation of 5\%, which
could be periodic, while 0856+172 shows periodic variability on a time 
scale of 40 minutes with a maximum peak-to-peak variation of
9\%. Further, both objects show a gradual decline in their mean level by 
$\sim 0.05$ and $\sim 0.15$ magnitude, respectively, during the period of our 
observations (2--3 hours). The magnitude of microvariability observed in these
two objects are slightly more than reported for the non-BAL radio-quiet sample
of Gopal-Krishna et al (1995) and Sagar et al (1996), which show variability 
of the order 3--5\%. The radio-quiet sample also do not show any gradual 
rise/decline of the magnitude levels as seen here.

The \ion{C}{4} BAL troughs fall within the $V$ band in 0846+156, while they fall
outside the $V$ band in 0856+172, ruling out the possibility of the observed
light curve variations being caused by variations in the strength of the BAL 
features. This result is similar to the conclusion of Barlow et al (1992), who
found the variation in the absorption equivalent widths in CSO 203 are not a 
result of the changes in the broad band fluxes which are $\le 10$\%.
The light curve variations observed in the two objects presented here are 
continuum variations as in the case of both the radio-quiet and the radio-loud 
objects. 

The observed variability may either be directly linked with the BAL phenomena, or, the
choice of the BAL QSO sample produces a selection of objects more conducive for
time variability studies. In the former case of the variability being caused by
the BAL phenomena, it is unclear how the absorbing clouds typically at a 
distance of $\sim 1$ pc from the central continuum source could produce 
variability on such a short time scale. Neither a smoothly flowing BAL wind
(Hamann, Korista \& Morris 1993) or clumpy BAL clouds (Turnshek 1988) or stellar contrails
from evolved stars in the nucleus (Scoville \& Norman 1995) can easily account
for the observed microvariability.
A more likely possibility appears to be that the intrinsic optical microvariability
in the BAL and the non-BAL radio-quiet QSOs is of the same origin, but their
observability is more optimised in the case of the BAL systems. Such a scenario
would, in principle, arise if the BAL clouds participate in supersonic
outflows from the polar regions of a blackhole + accretion disk system and are closely
aligned to the line of sight to the observer. In such a case, the disc and all
the associated variability phenomena on its surface such as flares, hot spots etc 
would be seen face on. Thus the BALs select the most advantageous orientation
for viewing the variability phenomena and can explain the observed difference
in the microvariability with respect to the parent radio-quiet population
as an `orientation effect.' 

A most intriguing aspect of the present study appears to be the similarity in the
observed light curves with those of the  the BL Lacertae objects (Miller \& Carini 1991; Wagner et al
1991; Carini 1991; Wagner \& Witzel 1995), which are flat spectrum radio-loud sources.
The BL Lacs show a similar decline in the mean light curve, with a suggestion of 
microvariability similar to the ones observed here for the BAL QSOs. 
For example: in the data presented by Carini (1991), OJ 287 faded in brightness 
by $\sim 0.08$ magnitude in $\sim 3.7$ hours on one occasion, in addition to
rapid variability. In addition to rapid variations, decline in the magnitude
by $\sim 0.1$ magnitude has been detected in both BL Lac and OQ 530 (Miller \&
Carini 1991). Recently Becker et al (1997) have detected a flat spectrum 
radio-loud BAL QSO along with two other low ionisation BAL quasars from a 
radio-selected quasar sample. The studies of the properties of the three 
objects suggest a trend of increasing radio luminosity with the amount of 
absorption to the quasar, leading them to suggest their objects could be 
transition objects evolving from radio-loud to radio-quiet BAL systems as the 
QSO emerges from an enshrouding material.  These findings seem to favour the 
orientation effect as an explanation for the observed differences in the light 
curves of the BAL QSOs and non-BAL QSOs.

The detection of microvariability in radio-quiet non-BAL QSOs, the detection of 
flat spectrum radio-loud BAL QSOs, together with the results presented here
open the possiblilty that the BAL QSO systems are a direct link between the 
radio-loud BL Lacs and the radio-quiet QSO samples. 

Krishan \& Wiita (1994) have considered detailed physics of a variety of plasma 
mechanisms that could give rise to variability in AGNs from time scales of hours 
to years. These fall broadly under MHD fluctuations, flares, and plasma-modulated
electromagnetic wave instabilities. Of these, the last class appears to be
best suited to describe the short time scale variations seen here.

More observations of a larger sample are required to be able to detect
similarities/dissimilarities among the different class of objects and also
ascertain the nature of the observed microvariabilities.

\section{Summary}

We present the first results from a pilot programme to monitor the 
short term optical variability in BAL QSO systems. 
Intra-night optical variations of $\sim 5-9 \%$ were detected on time scales
of $\sim$ one hour in QSOs 0846+156 and 0856+172. Further, the mean
magnitude level decreased in the two objects by $\sim 0.05$ and $\sim 0.15$
magnitude respectively during the period of our observations. The observed light 
curves are quite similar to those previously seen in the flat spectrum 
radio-loud sources, especially the BL Lacertae objects. QSO 0856+172 showed a
sudden dip in the light curve by $\sim 1$~mag.

\acknowledgements

We thank the VBT time allocation committee for allotment of time for this 
programme, and also thank the referee for useful comments.

\clearpage
\section*{Figure captions}

\figcaption{{\it Top:} $V$ band differential light curve for QSO 0846+156 with 
respect to comparison star ($c_1$).
{\it Bottom:} Difference between the comparison star ($c_1$) and the check star
($c_2$).} 

\figcaption{{\it Top:} $V$ band differential light curve for QSO 0856+172 with 
respect to comparison star ($c_1$). Note the 1 mag dip around UT 20.5.
{\it Bottom:} Difference between the comparison star ($c_1$) and the check star
($c_2$).} 

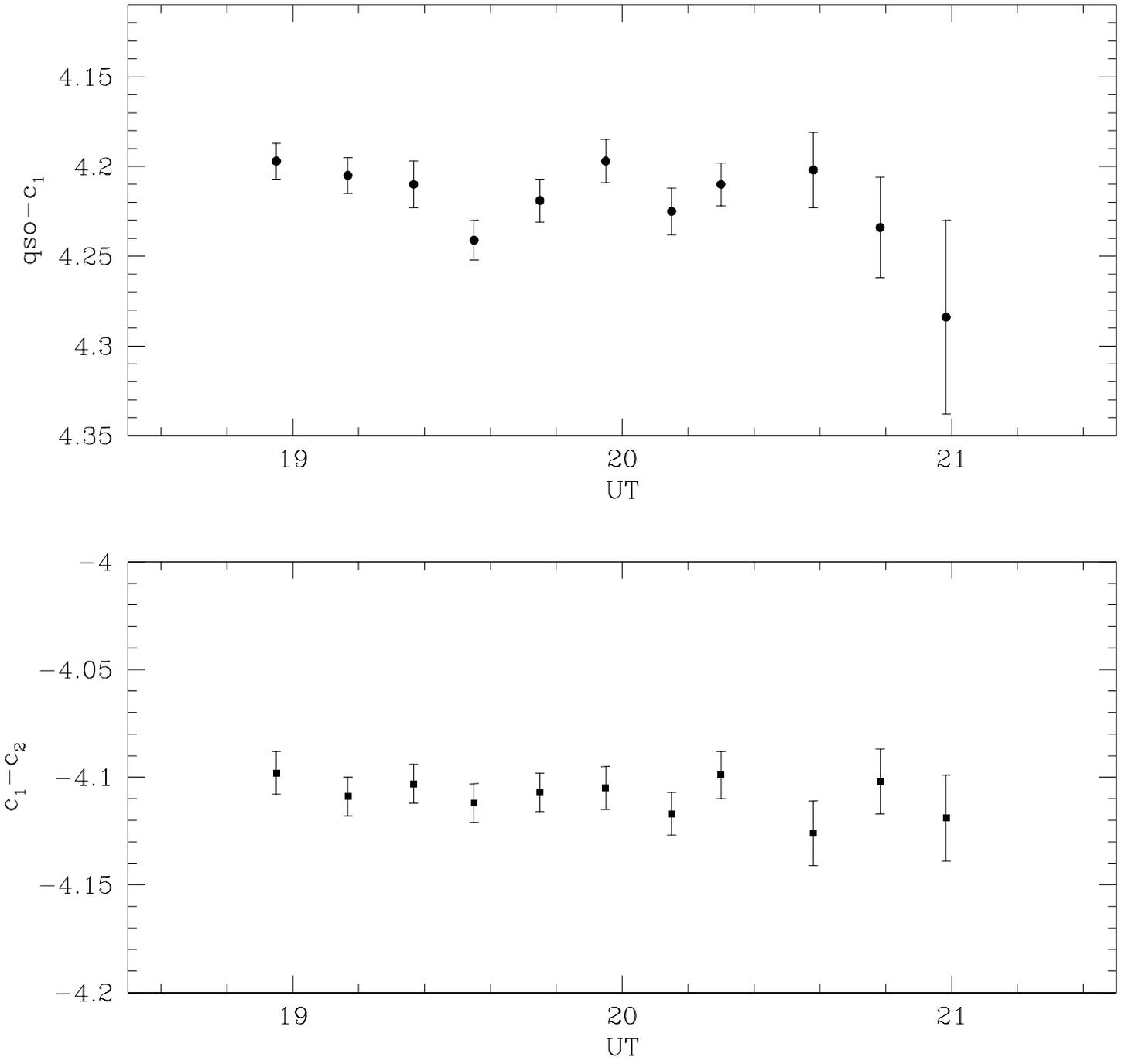
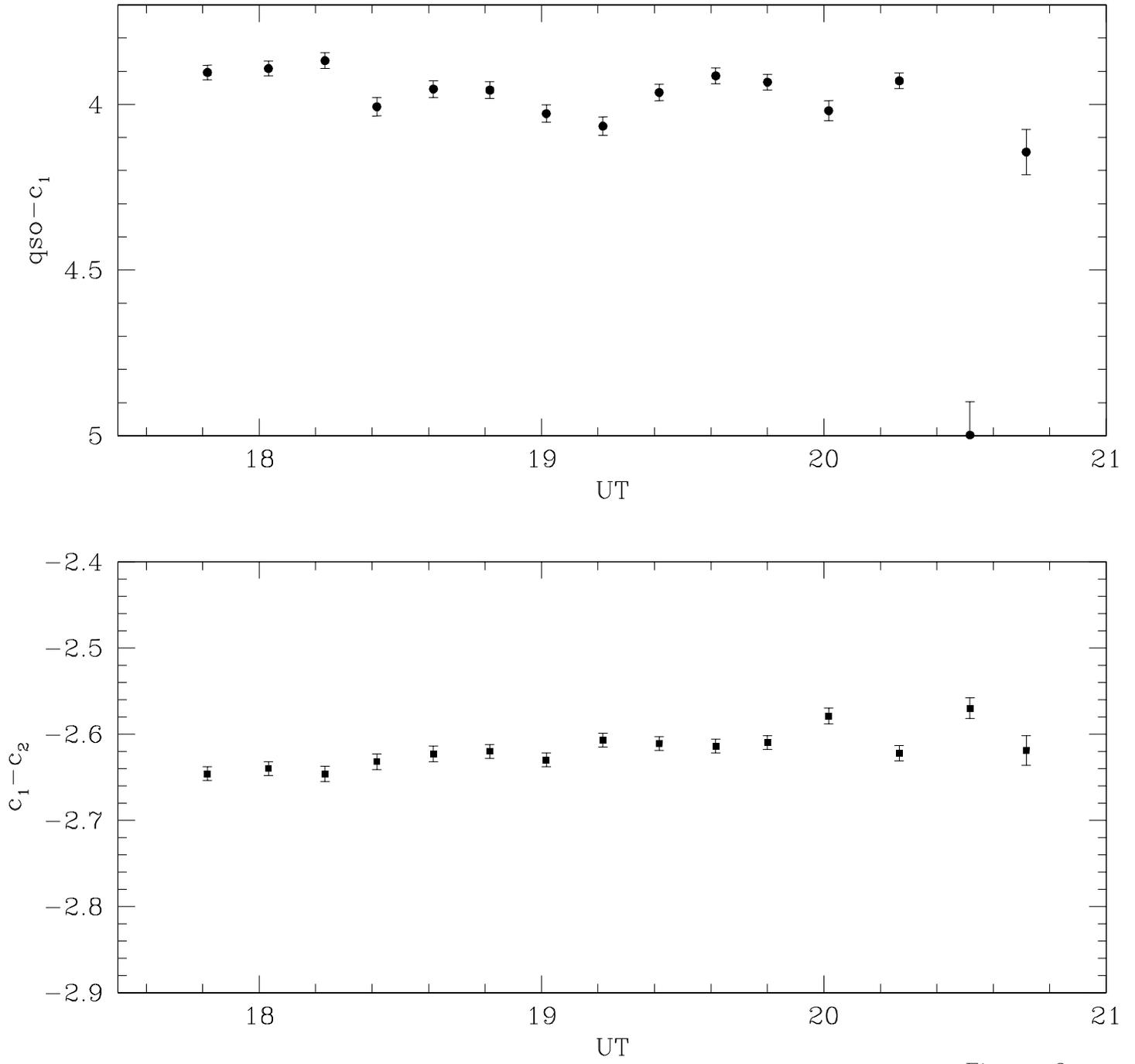
\end{document}